\documentclass[twocolumn]{aastex631}

\usepackage{comment}
\usepackage{CJK}

\newcommand{\kms}{\mbox{km s$^{-1}$}}
\newcommand{\x}{\mbox{$\times$}}
\newcommand{\Msun}{\mbox{$M_{\odot}$}}
\newcommand{\Lsun}{\mbox{$L_{\odot}$}}

\submitjournal{\apj}

\shorttitle{The Magnetic Field at the Onset of High-mass Star Formation}
\shortauthors{Sanhueza et al.}
\graphicspath{{./}{figures/}}

\begin{document}

\title{Magnetic Fields in Massive Star-forming Regions (MagMaR). V. The Magnetic Field at the Onset of High-mass Star Formation}

\correspondingauthor{Patricio Sanhueza}
\email{patosanhueza@gmail.com}

\author[0000-0002-7125-7685]{Patricio Sanhueza}
\affiliation{Department of Earth and Planetary Sciences, Institute of Science Tokyo, Meguro, Tokyo, 152-8551, Japan}
\affiliation{National Astronomical Observatory of Japan, National Institutes of Natural Sciences, 2-21-1 Osawa, Mitaka, Tokyo 181-8588, Japan}
\affiliation{Department of Astronomical Science, SOKENDAI (The Graduate University for Advanced Studies), 2-21-1 Osawa, Mitaka, Tokyo 181-8588, Japan}

\begin{CJK*}{UTF8}{gbsn}
\author[0000-0002-4774-2998]{Junhao Liu (刘峻豪)}
\affiliation{National Astronomical Observatory of Japan, National Institutes of Natural Sciences, 2-21-1 Osawa, Mitaka, Tokyo 181-8588, Japan}

\author[0000-0002-6752-6061]{Kaho Morii}
\affil{Department of Astronomy, Graduate School of Science, The University of Tokyo, 7-3-1 Hongo, Bunkyo-ku, Tokyo 113-0033, Japan}
\affil{National Astronomical Observatory of Japan, National Institutes of Natural Sciences, 2-21-1 Osawa, Mitaka, Tokyo 181-8588, Japan} 

\author[0000-0002-3829-5591]{Josep Miquel Girart}
\affiliation{Institut de Ciencies de l'Espai (ICE, CSIC), Can Magrans s/n, 08193, Cerdanyola del Vallès, Catalonia, Spain}
\affiliation{Institut d'Estudis Espacials de Catalunya (IEEC), 08034, Barcelona, Catalonia, Spain}

\author[0000-0003-2384-6589]{Qizhou Zhang}
\affiliation{Center for Astrophysics $|$ Harvard \& Smithsonian, 60 Garden Street, Cambridge, MA 02138, USA}

\author[0000-0003-3017-4418]{Ian W. Stephens} 
\affiliation{Department of Earth, Environment, and Physics, Worcester State University, Worcester, MA 01602, USA}

\author[0000-0002-3466-6164]{James M. Jackson}
\affiliation{Green Bank Observatory, 155 Observatory Rd., Green Bank WV 24944, USA}

\author[0000-0002-3583-780X]{Paulo C. Cort\'es}
\affiliation{Joint ALMA Observatory, Alonso de C\'ordova 3107, Vitacura, Santiago, Chile}
\affiliation{National Radio Astronomy Observatory, 520 Edgemont Road, Charlottesville, VA 22903, USA}

\author[0000-0003-2777-5861]{Patrick M. Koch}
\affiliation{Academia Sinica, Institute of Astronomy and Astrophysics, No.1, Sec. 4, Roosevelt Rd., Taipei, Taiwan}

\author[0000-0001-6725-1734]{Claudia J. Cyganowski}
\affiliation{SUPA, School of Physics and Astronomy, University of St. Andrews, North Haugh, St. Andrews KY16 9SS, UK}

\author[0000-0002-0028-1354]{Piyali Saha}
\affiliation{National Astronomical Observatory of Japan, National Institutes of Natural Sciences, 2-21-1 Osawa, Mitaka, Tokyo 181-8588, Japan}

\author[0000-0002-1700-090X]{Henrik Beuther}
\affiliation{Max Planck Institute for Astronomy, K\"onigstuhl 17, 69117 Heidelberg, Germany}

\author[0000-0002-8389-6695]{Suinan Zhang (张遂楠)}
\affiliation{Shanghai Astronomical Observatory, Chinese Academy of Sciences, 80 Nandan Road, Shanghai 200030, People’s Republic of China}

\author[0000-0003-3315-5626]{Maria T.\ Beltr\'an}
\affiliation{INAF-Osservatorio Astrofisico di Arcetri, Largo E. Fermi 5, I-50125 Firenze, Italy}

\author[0000-0002-8691-4588]{Yu Cheng}
\affiliation{National Astronomical Observatory of Japan, National Institutes of Natural Sciences, 2-21-1 Osawa, Mitaka, Tokyo 181-8588, Japan}

\author[0000-0002-8250-6827]{Fernando A. Olguin}
\affiliation{Yukawa Institute for Theoretical Physics, Kyoto University, Kyoto, 606-8502, Japan}
\affiliation{National Astronomical Observatory of Japan, National Institutes of Natural Sciences, 2-21-1 Osawa, Mitaka, Tokyo 181-8588, Japan}
\affiliation{Institute of Astronomy and Department of Physics,
National Tsing Hua University, Hsinchu 300044, Taiwan}

\author[0000-0003-2619-9305]{Xing Lu (吕行)}
\affiliation{Shanghai Astronomical Observatory, Chinese Academy of Sciences, 80 Nandan Road, Shanghai 200030, People’s Republic of China}

\author[0000-0002-7497-2713]{Spandan Choudhury}
\affiliation{Korea Astronomy and Space Science Institute (KASI), 776 Daedeokdae-ro, Yuseong-gu, Daejeon 34055, Republic of Korea}

\author[0000-0002-8557-3582]{Kate Pattle}
\affiliation{Department of Physics and Astronomy, University College London, Gower Street, London WC1E 6BT, United Kingdom}

\author[0000-0001-5811-0454]{Manuel Fern\'andez-L\'opez}
\affiliation{Instituto Argentino de Radioastronomía (CCT- La Plata, CONICET, CICPBA, UNLP), C.C. No. 5, 1894, Villa Elisa, Buenos Aires, Argentina}

\author[0000-0001-7866-2686]{Jihye Hwang}
\affil{Korea Astronomy and Space Science Institute (KASI), 776 Daedeokdae-ro, Yuseong-gu, Daejeon 34055, Republic of Korea}

\author[0000-0001-7379-6263]{Ji-hyun Kang}
\affiliation{Korea Astronomy and Space Science Institute (KASI), 776 Daedeokdae-ro, Yuseong-gu, Daejeon 34055, Republic of Korea}

\author[0000-0001-5996-3600]{Janik Karoly}
\affiliation{Department of Physics and Astronomy, University College London, Gower Street, London WC1E 6BT, United Kingdom}

\author[0000-0001-6431-9633]{Adam Ginsburg }
\affiliation{Department of Astronomy, University of Florida, P.O. Box 112055, Gainesville, FL 32611, USA}

\author[0000-0002-9907-8427]{A. -Ran Lyo}
\affiliation{Korea Astronomy and Space Science Institute (KASI), 776 Daedeokdae-ro, Yuseong-gu, Daejeon 34055, Republic of Korea}

\author[0000-0003-4402-6475]{Kotomi Taniguchi}
\affiliation{National Astronomical Observatory of Japan, National Institutes of Natural Sciences, 2-21-1 Osawa, Mitaka, Tokyo 181-8588, Japan}

\author[0000-0001-9822-7817]{Wenyu Jiao}
\affiliation{Shanghai Astronomical Observatory, Chinese Academy of Sciences, 80 Nandan Road, Shanghai 200030, People’s Republic of China}

\author[0000-0003-4761-6139]{Chakali Eswaraiah}
\affiliation{Department of Physics, Indian Institute of Science Education and Research (IISER) Tirupati, Yerpedu, Tirupati - 517619, Andhra Pradesh, India}

\author[0000-0003-4506-3171]{Qiu-yi Luo (罗秋怡)}
\affiliation{Shanghai Astronomical Observatory, Chinese Academy of Sciences, 80 Nandan Road, Shanghai 200030, People’s Republic of China}
\affiliation{School of Astronomy and Space Sciences, University of Chinese Academy of Sciences, No. 19A Yuquan Road, Beijing 100049, People’s Republic of China}
\affiliation{Key Laboratory of Radio Astronomy and Technology, Chinese Academy of Sciences, A20 Datun Road, Chaoyang District, Beijing, 100101, P. R. China}

\author[0000-0002-6668-974X]{Jia-Wei Wang}
\affiliation{Academia Sinica, Institute of Astronomy and Astrophysics, No.1, Sec. 4, Roosevelt Rd., Taipei, Taiwan}
\affiliation{East Asian Observatory, 660 N. A'oh\={o}k\={u} Place, University Park, Hilo, HI 96720, USA}

\author[0000-0003-2407-1025]{Beno\^it Commer\c{c}on}
\affiliation{Univ. Lyon, Ens de Lyon, Univ. Lyon 1, CNRS, Centre de Recherche Astrophysique de Lyon UMR5574, 69007, Lyon, France}

\author[0000-0003-1275-5251]{Shanghuo Li}
\affiliation{Max Planck Institute for Astronomy, K\"onigstuhl 17, 69117 Heidelberg, Germany}

\author[0000-0001-5950-1932]{Fengwei Xu}
\affiliation{Kavli Institute for Astronomy and Astrophysics, Peking University, Beijing 100871, People's Republic of China}
\affiliation{I. Physikalisches Institut, Universität zu Köln, Zülpicher Str. 77, D-50937 Köln, Germany}
\affiliation{Department of Astronomy, School of Physics, Peking University, Beijing, 100871, People's Republic of China}

\author[0000-0002-9774-1846]{Huei-Ru Vivien Chen}
\affiliation{Institute of Astronomy and Department of Physics,
National Tsing Hua University, Hsinchu 300044, Taiwan}

\author[0000-0003-2343-7937]{Luis A. Zapata}
\affil{Instituto de Radioastronom\'\i a y Astrof\'\i sica, Universidad Nacional Aut\'onoma de M\'exico, P.O. Box 3-72, 58090, Morelia, Michoac\'an, M\'exico}

\author[0000-0003-0014-1527]{Eun Jung Chung}
\affiliation{Korea Astronomy and Space Science Institute (KASI), 776 Daedeokdae-ro, Yuseong-gu, Daejeon 34055, Republic of Korea}

\author[0009-0007-6357-6874]{Fumitaka Nakamura}
\affil{National Astronomical Observatory of Japan, National Institutes of Natural Sciences, 2-21-1 Osawa, Mitaka, Tokyo 181-8588, Japan} 
\affil{Department of Astronomy, Graduate School of Science, The University of Tokyo, 7-3-1 Hongo, Bunkyo-ku, Tokyo 113-0033, Japan}

\author[0009-0007-6357-6874]{Sandhyarani Panigrahy}
\affiliation{Department of Physics, Indian Institute of Science Education and Research (IISER) Tirupati, Yerpedu, Tirupati - 517619, Andhra Pradesh, India}

\author[0000-0003-4521-7492]{Takeshi Sakai}
\affiliation{Graduate School of Informatics and Engineering, The University of Electro-Communications, Chofu, Tokyo 182-8585, Japan}



\begin{abstract}

A complete understanding of the initial conditions of high-mass star formation and what processes determine multiplicity require the study of the magnetic field in young, massive cores. Using ALMA 250 GHz polarization observations (0\farcs3 = 1000 au) and ALMA 220 GHz high-angular resolution observations (0\farcs05 = 160 au), we have performed a full energy analysis including the magnetic field at core scales and have assessed what influences the multiplicity inside a massive core previously believed to be in the prestellar phase. With a mass of 31 \Msun, the G11.92 MM2 core has a young CS molecular outflow with a dynamical time scale of a few thousand years. At high-resolution, the MM2 core fragments into a binary system with a projected separation of 505 au and a binary mass ratio of 1.14. Using the DCF method with an angle dispersion function analysis, we estimate in this core a magnetic field strength of 6.2 mG and a mass-to-magnetic flux ratio of 18. The MM2 core is strongly subvirialized with a virial parameter of 0.064, including the magnetic field. The high mass-to-magnetic flux ratio and low virial parameter indicate that this massive core is very likely undergoing runaway collapse, which is in direct contradiction with the core-accretion model. The MM2 core is embedded in a filament that has a velocity gradient consistent with infall. In line with clump-fed scenarios, the core can grow in mass at a rate of 1.9--5.6 \x\ 10$^{-4}$ \Msun\ yr$^{-1}$. In spite of the magnetic field having only a minor contribution to the total energy budget at core scales (a few 1000s au), it likely plays a more important role at smaller scales (a few 100s au) by setting the binary properties. Considering energy ratios and a fragmentation criterion at the core scale, the binary system could have been formed by core fragmentation. The binary system properties (projected separation and mass ratio), however, are also consistent with radiation-magnetohydrodynamic simulations with super-Alfvenic, supersonic (or sonic) turbulence that form binaries by disk fragmentation.

\end{abstract}



\section{Introduction} \label{sec:intro}
\end{CJK*}

How high-mass cores gather the necessary mass to form high-mass stars is still uncertain. Were they slowly formed under virial equilibrium conditions with most of their mass already in place early on, in a high-mass prestellar core? Or, did they start with low mass under subvirial conditions?

The search for high-mass prestellar cores has been intense \citep{Zhang09,Sanhueza13,Sanhueza17,Sanhueza19,Tan13,Wang14,Ohashi16,Contreras18,Pillai19,Li19b,Morii21,Redaelli22}, resulting in no detections at the early evolutionary stages found in infrared dark clouds \citep[IRDCs;][]{Rathborne06,Chambers09,Sanhueza12}. Among the recent studies of 70 $\mu$m dark IRDCs, the ASHES survey \citep[The ALMA Survey of 70 $\mu$m Dark High-mass Clumps in Early Stages;][]{Sanhueza19} shows a complete absence of high-mass prestellar cores ($>$30 \Msun) in 39 massive clumps containing 839 cores \citep{Morii23,Morii24}. The core dynamics analysis of the ASHES pilot survey \citep{Li23} shows that at larger masses, both prestellar and protostellar cores are more subvirialized, i.e., have lower virial parameters ($\alpha = M_{\rm vir}/M<$ 1, with $M_{\rm vir}$ being the virial mass and $M$ the total gas mass). Considering only turbulence and gravity, more massive cores are farther out of equilibrium and likely to collapse fast, unless there are other energies at play that could counter gravity, such as the magnetic field and rotation. Indeed, using ALMA, the magnetic field has been mapped in IRDC G28.34+0.06 \citep{Liu20} and IRDC 18310-4 \citep{Beuther18,Morii21}, finding that cores can have larger virial parameters after including the magnetic field, but the most massive cores remain subvirial. 

The rare high-mass prestellar core candidates found so far tend to be embedded in clumps more evolved than 70 $\mu$m dark IRDCs \citep[e.g.,][]{Tan13,Cyganowski14,Liu17,Nony18,Barnes23,Mai24}. 
Thorough analysis of their kinematics, physical, and/or chemical properties, however, found consistency with protostellar activity \citep[e.g.,][]{Feng16,Tan16,Molet19,Cyganowski22}. Nevertheless, the high-mass cores previously believed to be in the prestellar phase offer the opportunity to study a high-mass core at the onset of star formation. 

In this regard, the massive core G11.92 MM2 with a mass $>$30 \Msun\ \citep{Cyganowski14} is sufficiently bright to be observed in mm/submm polarization, allowing the study of the magnetic field and its importance in the star formation process. Because the evidence of star formation in G11.92 MM2 suggests that it is very young, the energy balance and virial equilibrium analysis in this young protostellar core can be used as a proxy to infer which physical conditions were present in the prestellar phase. Another key aspect for which the study of a young core is of great advantage is for determining which physical processes influence  multiplicity. The fragmentation of cores and/or accretion disks is regulated not only by gravity and thermal pressure, but also by turbulence and magnetic fields that can introduce asymmetries  and rotation, promoting the formation of multiple systems \citep{Offner23}. On this point, the recently discovered binary system in G11.92 MM2 \citep{Cyganowski22} offers the possibility to study which physical conditions caused the fragmentation. 

The G11.92 region, located at a distance of 3.37 kpc \citep{Sato14}, is forming a young stellar cluster \citep{Cyganowski17}. The brightest object in the region is MM1, a high-mass star of $\sim$1.2 \x\ 10$^4$ \Lsun\ \citep{Moscadelli16} forming through an accretion disk \citep{Ilee16,Ilee18}. While the MM2 core eluded the detection of signs indicating active star formation for almost a decade, the finding of an embedded proto-binary system with 1.3 mm brightness temperatures indicative of internal heating and a low-velocity molecular outflow in CH$_3$OH emission reveal star formation activity at 100s au scales. More recently, \cite{Zhang24} show that the MM2 core is actively accreting mass from its large-scale environment. \cite{Cyganowski22} suggest that the binary may be forming in a weakly magnetized environment, which could be confirmed by full polarization observations that can unveil the local magnetic field. 

In the Magnetic fields in Massive star-forming Regions (MagMaR) project, we have observed 30 high-mass star-forming regions in polarization at 1.2 mm ($\sim$250 GHz) using ALMA. The first papers describing results from this project  show a variety of magnetic field morphologies: radially oriented due to an explosive event \citep{Fernandez21}, hourglass-like implying a strong field \citep{Cortes21,Saha24}, spiral-shaped indicating a gravity-dominated system \citep{Sanhueza21,Cortes24}, and aligned with velocity gradients that likely trace material flowing toward a high-mass star \citep{Zapata24}. As part of MagMaR, we have observed the G11.92 high-mass star-forming region, including the aforementioned former high-mass prestellar core MM2. While the MagMaR project offers mapping the magnetic field at core scales, the Digging into the Interior of Hot Cores with ALMA (DIHCA) survey provides a view on how the cores embedded in the same 30 fields fragment at a few 100 au scales using ALMA at 220 GHz \citep{Olguin22,Olguin23,Taniguchi23,Li24,Ishihara24}. 

In this paper, we combine observations of G11.92 MM2 from the MagMaR and DIHCA surveys to measure the core-scale magnetic field and constrain the importance of the magnetic fields at the earliest stages of massive binary formation. 

\section{Observations} \label{sec:obs}

\subsection{Polarization Observations}

\begin{figure*}[ht!]
\epsscale{1.17}
\plotone{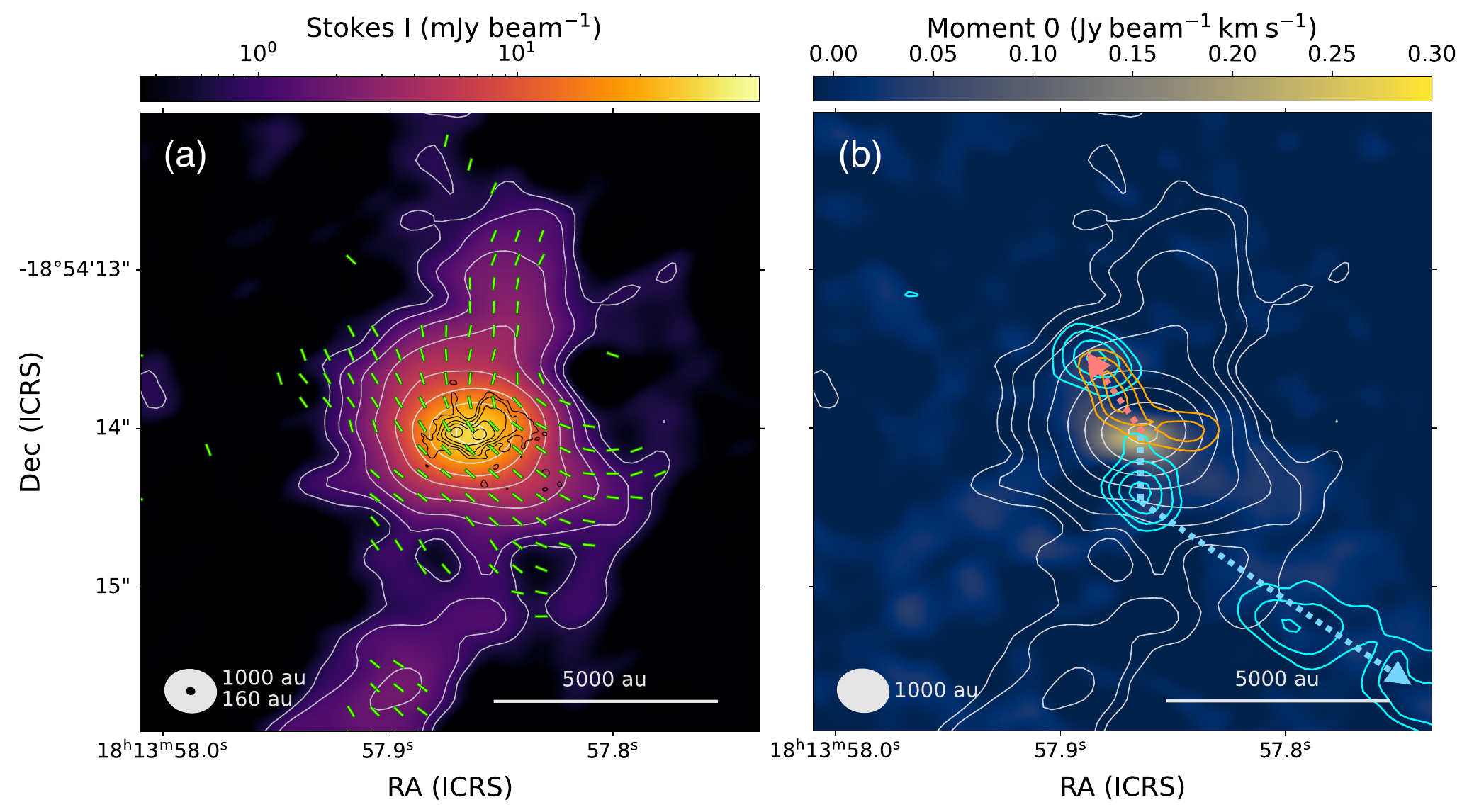}
\caption{(a) ALMA 1.2 mm dust continuum emission (color scale and white contours) toward G11.92-0.61 MM2 with overlaid magnetic field segments. Green line segments representing the magnetic field orientation (dust polarization vectors rotated by 90 deg, plotted in Nyquist sampling) are plotted above the 3$\sigma_{QU}$ level, with $\sigma_{QU}=30$ $\mu$Jy beam$^{-1}$, and have an arbitrary length. White contours correspond to the dust continuum emission at low resolution ($\sim$0\farcs3) in steps of 4, 6, 10, 18, 34, 66, 130, and 258 times the $\sigma_I$ (rms) value of 181 $\mu$Jy beam$^{-1}$. Black contours correspond to the 1.33 mm dust continuum emission at high-angular resolution  ($\sim$0\farcs05) in steps of 5, 10, 20, 40, and 80 times the $\sigma$ value of 36.1 $\mu$Jy beam$^{-1}$. (b) Moment 0 map of the $K=0$ and $K=1$ transitions combined of CH$_3$CN $(v = 0, J = 14 -13)$ in color scale. 
White contours represent the dust continuum emission as in panel (a). Cyan and orange contours correspond to the blueshifted and redshifted outflow emission traced by CS $J=5-4$ in steps of 3, 5, 8, and 11 times the $\sigma$ value of 10 mJy beam$^{-1}$ \kms\ and in steps of 3, 4, 5, and 6 times the $\sigma$ value of 4.7 mJy beam$^{-1}$ \kms, respectively. The blueshifted component is integrated from 18 to 32.9 \kms, and the redshifted component from 41.3 to 44.9 \kms. The cyan and orange arrows represent the direction of the CH$_3$OH outflow detected by \cite{Cyganowski22}. The white and black beams at the bottom left represent the spatial resolution of the polarization observations at 1000 au (0\farcs3) and the high-angular resolution observations at 160 au (0\farcs05), respectively. 
 Scale bar is shown on the bottom, right side of each panel.
\label{Bfield}}
\end{figure*}

Full polarization observations were taken as part of the MagMaR survey (Project ID: 2017.1.00101.S and 2018.1.00105.S; PI: Sanhueza). Observations at 1.2 mm ($\sim$250.486 GHz) of G11.92-0.61 were carried out on September 25, 2018 using 47 antennas of the 12 m array. An angular resolution of $\sim$0\farcs3 ($\sim$1000 au) was obtained with unprojected baselines ranging from 15 to 1400 m. The correlator setup includes five spectral windows: two spectral windows of 234 MHz width, providing a spectral resolution of 0.488 MHz (0.56 \kms), and three spectral windows of 1875 MHz width, with a spectral resolution of 1.95 MHz (2.4 \kms). Data calibration and imaging were performed using CASA versions 5.1.1 and 5.5.0,  respectively.

We adopted the procedures described in \cite{Olguin21} to remove channels with line emission from the continuum (Stokes {\it I}) image. Stokes {\it I} was self-calibrated in phase and amplitude. Self-calibration solutions were also applied to the spectral cubes. 

Each Stokes parameter image was independently cleaned using the CASA task {\it tclean} with Briggs weighting and robust parameter equal to 1. The final images have an angular resolution of 0\farcs27 \x\ 0\farcs34 and sensitivities of 181 $\mu$Jy beam$^{-1}$ for Stokes {\it I} ($\sigma_{I}$), and 30.3 $\mu$Jy beam$^{-1}$ for both Stokes {\it Q} and {\it U} ($\sigma_{QU}$). The polarized intensity image was debiased following \cite{Vaillancourt06}. 

 The automatic masking procedure {\it yclean} presented in \cite{Contreras18} was used for the imaging of the line emission. 
The CASA task {\it tclean} with Briggs weighting and a robust parameter equal to 1 was used, producing a noise level of 1.3 mJy beam$^{-1}$ per $\sim$2.4 \kms\ channel for CS $J=5-4$ and CH$_3$CN $(v = 0,\,J = 14 -13)$, and 3.0 mJy beam$^{-1}$ per 0.56 \kms\ channel for H$^{13}$CO$^+$ $J=3-2$.  

\subsection{High-resolution Observations}

Long baseline observations were taken as part of the DIHCA survey (Project ID: 2016.1.01036.S and 2017.1.00237.S; PI: Sanhueza). Observations at 1.33 mm ($\sim$226.2 GHz) of G11.92-0.61 were carried out on September 11, 2017 and July 28, 2019, using 42 antennas of the 12 m array. An angular resolution of $\sim$0\farcs048 ($\sim$160 au) was obtained with unprojected baselines ranging from 42 to 8548 m. The correlator setup includes four spectral windows of 1875 MHz width and a spectral resolution of 976 kHz ($\sim$1.3 \kms). Data calibration and imaging were performed using CASA versions 5.6.1-8 and 5.7.0,  respectively. 

Like the polarization data, we also remove channels with line emission from the continuum via the procedure in \cite{Olguin21}. In the DIHCA data, a large number of molecular lines have been detected in the core MM1 of G11.92, but none have been detected in the core MM2. Continuum emission was self-calibrated in phase and amplitude. The continuum emission was cleaned using the CASA task {\it tclean} with Briggs weighting and a robust parameter equal to 0.5. The final images have an angular resolution of 0\farcs055 \x\ 0\farcs042 and sensitivity of 36.1 $\mu$Jy beam$^{-1}$.  

\begin{figure*}[ht!]
\epsscale{1.2}
\plotone{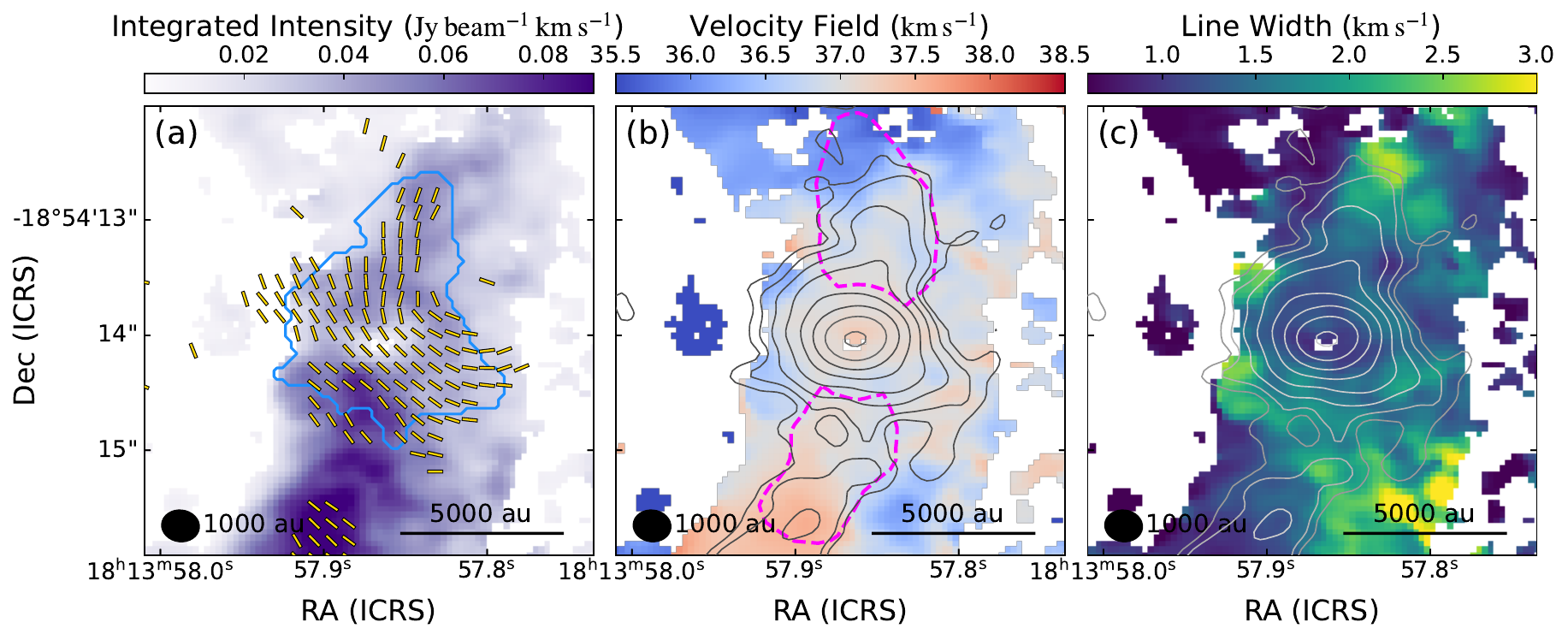}
\caption{Integrated intensity (a), velocity field (b), and line FWHM  (c) obtained from fitting a Gaussian component to the H$^{13}$CO$^+$ emission pixel-by-pixel. Contours show the dust continuum emission (same as in Figure~\ref{Bfield}). All pixels that have a peak intensity, as derived from the Gaussian fitting, larger than 4$\sigma$ ($\sigma$ =  3 mJy beam$^{-1}$) are displayed. In panel (a), the dendrogram leaf, defined from the continuum emission, is displayed in blue (see details in Appendix~\ref{sec:dust-app}) and the magnetic field vectors in yellow. Core properties were measured inside the leaf. In panel (b), the areas used to derive the properties of the infalling gas are shown in dashed, magenta masks. 
\label{fitting}}
\end{figure*}

\section{Results} \label{sec:res}

Figure~\ref{Bfield}a shows the 1.2 mm dust continuum emission image (and white contours) with the magnetic field directions projected on the plane of the sky in green segments at 1000 au scales. The centrally condensed MM2 core has dust emission extending to the northern and western parts that are followed by the magnetic field. At the center of the MM2 core, the 1.3 mm high-angular resolution data (160 au resolution, in black contours) from the DIHCA survey show the fragmentation of the MM2 core into a binary system \citep[MM2-E and MM2-W;][]{Cyganowski22}. 

So far, there has been no successful detection of compact molecular line emission emanating from the center of the MM2 core. In Figure~\ref{Bfield}b, we show compact CH$_3$CN ($v = 0$, $J = 14 -13$) emission (moment 0 map of the $K=0$ and $K=1$ transitions combined, $E_u/\kappa_{\rm B}$ equal to 92.7 and 99.8 K, respectively) coming from what seems to be a common compact structure hosting MM2-E and MM2-W. The detection of these transitions with $E_u/\kappa_{\rm B}$ $>$ 90 K implies internal heating and deeply embedded star formation activity. Figure~\ref{Bfield}b also shows a molecular outflow in CS $(J = 5 - 4)$. Considering the spatial resolution of 1000 au of the CS emission, it is unclear which of the binary members drives the outflow.  

Using CS outflow emission, the projected lengths of the blue- and red-shifted lobes ($l_{\rm b},\,l_{\rm r}$) are estimated to be 18000 and 2500 au (blueshifted emission toward the south-west and redshifted emission toward the north), respectively, while the maximum velocities 
($v_{\rm max\,b}=|v^{b}_{\rm LSR} - v_{\rm sys}|$ and $v_{\rm max\,r}=|v^{r}_{\rm LSR} - v_{\rm sys}|$ with $v_{\rm sys}=37.1$ \kms) correspond to 19.1 and 8.4 \kms, respectively. These values result in outflow dynamical timescales ($t_{\rm dyn} = l_{\rm b,r}/v_{\rm max\,b,r}$) of 4500 yr for the blueshifted lobe and 1400 yr for the redshifted lobe, supporting the idea that the MM2 core has recently entered the protostellar stage and it is at the onset of star formation.\footnote{The dynamical time scale has not been corrected by the unknown inclination of the outflow. \cite{Li20} calculate that for a mean inclination angle of $\sim$57.3\arcdeg, a correction factor of 0.6 should be applied to $t_{\rm dyn}$, resulting in dynamical ages of 2700 yr for the blueshifted lobe and 840 yr for the redshifted lobe.}

In addition to the compact and outflow emission, we also find in H$^{13}$CO$^+$ $(J = 3 - 2)$ emission a more extended gas component tracing the whole MM2 core and a larger filamentary structure connecting with the core from the north and south. Figure~\ref{fitting}a shows the integrated intensity of the single component of H$^{13}$CO$^+$ emission that has a spatial distribution coincident with that of the dust emission. 
The H$^{13}$CO$^+$ profiles are Gaussian-like, except at the center of the core, near the position of the CH$_3$CN emission, where the line emission exhibits a profile with absorption features at the velocity of the G11.92 cloud, i.e, 35.2 \kms\ \citep{Csengeri16}. Because the $J = 3 - 2$ transition presents a simple line profile and traces dense, relatively cold gas ($E_u/\kappa_{\rm B}$ = 25 K), it can be used to extract the kinematics of the core, namely rotation and turbulence. We have performed a Gaussian fitting, pixel-by-pixel, and the results are presented in Figure~\ref{fitting}b and c. Masked pixels at the center of the core correspond to places at which the Gaussian fitting failed. The integrated intensity map (Figure~\ref{fitting}a) shows weaker emission at the center of the core. 
We find no evidence of core rotation, meaning that the core, if rotating, could be doing so in or near the plane of the sky and/or the rotational velocity is small and unresolved at the current spectral resolution. This lack of rotation is not unusual at the early stages of high-mass star formation
\citep[e.g., cores named W43-MM1\#6 and G028.37+00.07 C2c1,][respectively]{Cunningham23,Barnes23}, and it can even occur in more evolved hot cores, for instance, \cite{Silva17,Saha24}. We do see, however, an overall velocity gradient in the filamentary structure that hosts the MM2 core, with a bluer component toward the north and a redder component toward the south (Figure~\ref{fitting}b). 

\section{Discussion} \label{sec:dis}

\subsection{MM2 Core Properties from Dust (Stokes I) and Line Emission}

To measure the properties of the MM2 core, we use the dendrogram definition made for the whole MagMaR core catalog presented in Sanhueza et al. (2025, in prep). The dendrogram leaf representing the core is shown in Figure~\ref{fitting}a (see Appendix~\ref{sec:dust-app} for details on dendrogram parameters). This leaf defines the area over which the velocity dispersion, magnetic field strength, and energies are later derived. The measured flux density and radius for the MM2 core at 1.2 mm are 163 mJy and 0\farcs37 (corresponding to 1250 au), respectively. Assuming optically thin dust emission and a dust temperature of 20 K \citep{Cyganowski14}, we obtain a mass and an average number density for the MM2 core of 31 $\pm$ 13 \Msun\ and 4.8 $\pm$ 2.5 $\times$ 10$^8$ cm$^{-3}$, respectively (more details can be found in Appendix~\ref{sec:dust-app}). 

The measured velocity width (FWHM) of the H$^{13}$CO$^+$ line in the leaf area defining MM2 is 1.44 \kms, corresponding to a Mach number ($\mathcal{M}$) of 2.3 (see details in Appendix~\ref{sec:kin}). This value of $\mathcal{M}$ is consistent with values found for the most massive cores embedded in IRDCs \citep{Li23}.

Using the high-angular resolution DIHCA data, we fitted a 2D Gaussian profile to both of the binary dust condensations to obtain their positions and fluxes. 
The projected separation between the two peak positions of 0\farcs15 (505 au) is consistent with the one derived by \cite{Cyganowski22}. The flux densities at 1.33 mm are 14.88 mJy for MM2-E and 16.91 mJy for MM2-W. Assuming that both condensations have the same dust temperature, their mass ratio is 1.14. 

\begin{figure}[ht!]
\epsscale{1.2}
\plotone{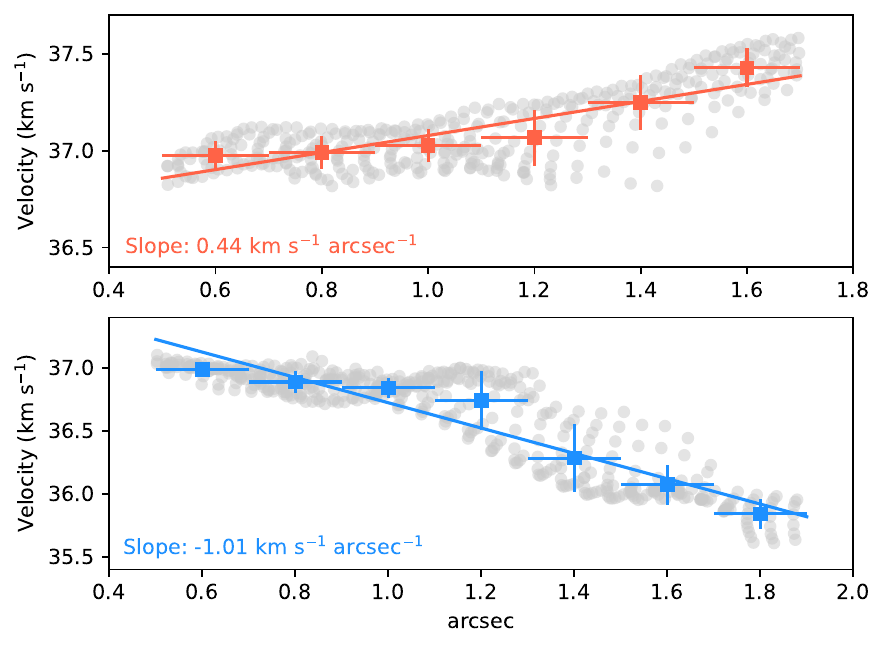}
\caption{Fitted velocity field to derive the velocity gradient in the filaments. The error bars correspond to the bin width and the standard deviation of the velocities inside each bin for the horizontal and vertical axes, respectively. In the top panel, the velocity gradient of 0.44 \kms\ arcsec$^{-1}$ corresponds to 26.9 \kms\ pc$^{-1}$, while in the bottom panel the velocity gradient of 1.01 \kms\ arcsec$^{-1}$ corresponds to 61.5 \kms\ pc$^{-1}$.
\label{filament_field}}
\end{figure}

\vspace{1cm}

\subsection{Mass feeding}

In recent years, it has become more frequent to find velocity gradients consistent with accretion flows (or streamers) in high-mass star-forming regions \citep[e.g.,][]{Peretto14,Liu15,Izquierdo18,Chen19,Sanhueza21,Olguin23,Fernandez23,Xu23,Alvarez24,Wells24}. The velocity field of the H$^{13}$CO$^+$ emission (Figure~\ref{fitting}b) shows how the northern gas becomes more blue-shifted away from the core, while the southern gas becomes more red-shifted away from the core. The v$_{\rm lsr}$ velocity around the MM2 core is $\sim$37.1 \kms. This pattern in the velocity gradient has previously been interpreted as flows of gas moving toward the center \citep[e.g.,][]{Kirk13,Peretto14}. We have fitted the velocity field inside an area approximately located at the 4$\sigma$ contour in the continuum (see the exact region in Figure~\ref{fitting}b). The velocity gradient ($\nabla V_{obs}$) was determined using a linear regression applied to average projected velocities in bins of 0\farcs2 with respect to the central source, similar to a radial velocity profile inside the masked area. 
The derived velocity gradient is of 61.5 \kms\ pc$^{-1}$ in the northern blue-shifted component and 26.9 \kms\ pc$^{-1}$ in the southern red-shifted component (Figure~\ref{filament_field}). Using the dust continuum emission and assuming a dust temperature of 20 K, the gas mass in the same area where the velocity gradients were measured is 4.0 and 2.5 \Msun\ for the blue-shifted and red-shifted flows, respectively. Following \cite{Kirk13} and using the measured $\nabla V_{obs}$, we can estimate the infall rate as $\dot M = M\, \nabla V_{obs} / \tan(i)$, resulting in 
$\dot M_{\rm blue} = 2.5 \times 10^{-4} / \tan(i)$ and $\dot M_{\rm red} = 6.9 \times 10^{-5} / \tan(i)$ \Msun\ yr$^{-1}$, with $i$ being the inclination angle. We derive a total infall rate of 1.9--5.6 \x\ $10^{-4}$ \Msun\ yr$^{-1}$ for arbitrary inclination angles between 60\arcdeg\ and 30\arcdeg, respectively, consistent with values derived in the region at larger scales using N$_2$H$^+$ emission by \cite{Zhang24}. 

\subsection{The Magnetic Field and Energy Balance} \label{sec:modeling}

To assess the importance of the magnetic field with respect to the other energies in play and to determine if the magnetic field has a role in the formation of the binary system, we estimate the magnetic field strength using the Davis-Chandrasekhar-Fermi method \citep[DCF;][]{Davis51,CF53} using the angle dispersion function approch \citep[ADF;][]{Houde09,Houde16}. As described in Appendix~\ref{sec:bfield-app1}, we estimate a magnetic field strength of 6.2 $\pm$ 3.5 mG, a mass-to-magnetic-flux ratio ($\lambda$) of 18 $\pm$ 9, and an Alfvenic Mach number ($\mathcal{M_{\rm A}}$) of 1.6 $\pm$ 0.7. In spite of having a relatively large field strength, the core is  unquestionably magnetically supercritical even with the uncertainties given by the DCF technique \citep{Liu21}.  The magnetic field cannot prevent the collapse of the core. 

In one of the scenarios that aims to explain the formation of high-mass stars, the core accretion theory, virial equilibrium of high-mass cores is important to enable a ``slow" collapse \citep[i.e., over several free-fall times;][]{Tan14}. Due to the general difficulty in evaluating the magnetic field strength in star-forming regions, virial analyses are frequently performed neglecting the magnetic field \citep[e.g.,][]{Li23}. Only considering the kinetic support, we calculate a virial parameter ($\alpha_{\rm vir}$) of 0.060 $\pm$ 0.026, far from equilibrium, $\alpha_{\rm vir}=1$ (Appendix~\ref{sec:kin}). However, as suggested in previous works  \citep[e.g.,][]{Tan13,Sanhueza17,Liu22a}, the magnetic field can add additional support to maintain a core in equilibrium. Following Appendix~\ref{sec:bfield-app1}, we find that the virial parameter including the magnetic energy ($\alpha_{\rm vir, B}$) is 0.064 $\pm$ 0.028. This value is consistent with the kinetic virial parameter, within the uncertainties, and it is clearly insufficient to bring the MM2 core to equilibrium. 

Lacking compact emission from {\it Spitzer}, the MM2 core was scrutinized for signs of star formation activity for many years using several observational facilities \cite[SMA, VLA, and ALMA;][]{Cyganowski14,Cyganowski17}. MM2 was considered one of the most promising high-mass prestellar candidates until the discovery of a weak outflow in CH$_3$OH \citep{Cyganowski22}, now also detected here in CS $J=5-4$. Indeed, the presence of only weak, compact CH$_3$CN emission and an outflow dynamical timescale of only a few thousand years strongly suggest that star formation must have been relatively recent. We can therefore adopt MM2 as a proxy to infer what the physical conditions were in the prestellar phase. Adopting the core accretion paradigm, we argue that extrapolating the core properties into the past prestellar phase, the virial parameter estimated at the current time is comparable or higher. First, the energies that can oppose collapse were likely weaker in the prestellar phase because the core should have been less turbulent than at present time ( e.g., no injection of turbulence into the envelope from a protostar). Second, the magnetic field should have been weaker because of the lower gas density  on average. However, under the assumption of core accretion, the core mass should have not significantly changed, making the gravitational energy approximately the same as the current measured value. We therefore conclude that even with the inclusion of the magnetic field, high-mass stars cannot form in MM2 under equilibrium conditions. This conclusion stands in stark contradiction to the core accretion scenario.

Considering the environment in which MM2 is embedded, a filament with a velocity gradient consistent with a mass infall rate of a few times 10$^{-4}$ \Msun\ yr$^{-1}$, and a runaway collapse as indicated by the high mass-to-magnetic-flux ratio ($\lambda$ = 18 $\pm$ 9) and low virial parameter ($\alpha$ = 0.064 $\pm$ 0.028), clump-fed scenarios seem more likely. 

Using numerical simulations, \cite{Liu21} show that when the magnetic field is weak with respect to other energies, i.e., not in equipartition, the DCF with the ADF method overestimates the magnetic field strength. Therefore, the virial parameter including the magnetic field can be considered an upper limit. To consider the effect of the dust temperature in the core properties, we have calculated all core parameters at 50 K (see Table~\ref{tab:no_B} and \ref{tab:B} in the Appendix).  With a $\alpha_{\rm vir, B}$ equal to $0.23\pm0.10$ and $\lambda=10\pm5$, all conclusions hold. 

\subsection{Magnetic Field and Binary Formation}\label{sec:binary}

With a separation in the plane of the sky of 505 au and a mass ratio of 1.14, the binary system in MM2 is likely one of the youngest observed among the few multiple systems known in high-mass star-forming regions \citep{Zhang19,Guzman20,Tanaka20,Beltran21,Olguin22,Li24}. With the closest neighboring condensation at 2\farcs9 (almost 10000 au) away from the binary system \citep[embedded in the MM5 core;][]{Cyganowski17}, the binary system has been likely dynamically unperturbed and  its properties should reflect the physical conditions near the moment of its formation.

\cite{Tsuribe99} formalized a criterion to assess whether a core is prone to fragmentation that depends on the ratio of the thermal to gravitational energy ($\alpha'$) and the ratio of the rotational to gravitational energy ($\beta'$). Fragmentation requires the following two conditions be met simultaneously: $\alpha' \beta' < 0.12$ and $\alpha' < 0.5$. For the MM2 core, we obtain $\alpha' \beta' = 8 \pm 7 \times 10^{-6}$ and $\alpha' = 5 \pm 2 \times 10^{-3}$ (see details in Appendix~\ref{sec:kin}).  The formalism derived by \cite{Tsuribe99}, which ignores the magnetic field, indicates that the binary in the MM2 core can be formed through core fragmentation. Although not dominant, the presence of the magnetic field can nevertheless play against core fragmentation \citep{Commercon11,Myers13}. 
\cite{Machida08} find for models that have initial ratios of $E_{\rm rot}/E_{\rm G}$ equal to 0.005, comparable to the values in the MM2 core (0.002-0.005 for 20-50 K), core fragmentation occurs for $\lambda > 10$ (consistent with values in the MM2 core, $\lambda$ of 10-18 for 20-50 K). The combination of absent/weak rotation at the core scale and the presence of a non-energetically dominant magnetic field, along with an overall dominant gravitational energy, makes the formation of the binary system through core fragmentation viable in the MM2 core.

There are few simulations of cores that end in forming binaries including high-mass stars \citep[e.g.,][]{Krumholz09,Mignon21,Mignon23}. In radiation-hydrodynamical  simulations, \cite{Krumholz09} obtain a binary system formed by disk fragmentation with a mass ratio of 1.4 and a separation of 1590 au. \cite{Mignon23} argue that the magnetic field can remove angular momentum in the innermost core regions, contributing to having smaller disks and smaller binary separations than hydrodynamical simulations suggest. \cite{Mignon21} report the formation of binary systems from disk fragmentation, rather than core fragmentation, when turbulence dominates over the magnetic field (super-Alfvenic turbulence). This is in agreement with our observations of MM2, with an Alfvenic Mach number ($\mathcal{M_{\rm A}}$) of 1.6 $\pm$ 0.7. 

Among available radiation-magnetohydrodynamical simulations of cores forming binary systems including high-mass stars, the so-called SUPAS simulation run with super-Alfvenic, supersonic turbulence from \cite{Mignon21}, results in the best match with the properties derived from our observations. Some of the conditions of the SUPAS simulation include a sonic Mach number ($\mathcal{M}$), an Alfvenic Mach number ($\mathcal{M_{\rm A}}$), and a mass-to-magnetic flux ratio normalized to the critical value ($\lambda$) of 2, 5.7, and 5, respectively, while the measured values toward MM2 correspond to 2.30 $\pm$ 0.03, 1.6 $\pm$ 0.7, 18 $\pm$ 9, respectively. However, the SUPA simulation with super-Alfvenic, sonic turbulence cannot be fully discarded considering that the turbulence in the prestellar phase could have been lower ($\mathcal{M}=0.5$, $\mathcal{M_{\rm A}}=1.4$, $\lambda = 5$). There are differences between the observations and the simulation conditions, however. For example, the magnetic field seems to be less important with respect to turbulence for the SUPAS simulation ($\mathcal{M_{\rm A}}=5.7$) and comparable for the SUPA simulation ($\mathcal{M_{\rm A}}=1.4$), but the magnetic field is more important with respect to gravity in both simulations ($\lambda = 5$). In spite of the differences, the obtained binary system formed through disk fragmentation has mass ratios between 1.1 and 1.6, and separations that range from 400 to 700 au for SUPAS and 350 to 600 au for SUPA, in good agreement with the observations. 

Whether the binary system embedded in MM2 formed through core fragmentation or disk fragmentation, the magnetic field likely influences the fragmentation process at a few 100s au scales. In the case of core fragmentation, it may reduce the number of fragments, while in disk fragmentation, it could determine the binary's properties, such as mass ratio and separation. Now that observations offer important constraints to theoretical models, numerical simulations with clear, testable predictions are essential for distinguishing between these two fragmentation modes. 

\section{Conclusions} \label{sec:conclu}

Studies including a complete energy analysis at the earliest stages of high-mass star formation are rare, especially at $\leq$1000 au scales. Numerical simulations have tackled the origin of multiplicity by exploring the influence of different physical processes. Observations to test the validity of the resulting theoretical simulations are frequently difficult to obtain, for example, polarization and long baseline interferometric observations. Combining the MagMaR (Magnetic fields in Massive star-forming Regions) and the DIHCA (Digging into the Interior of Hot Cores with ALMA) surveys, we have analyzed the magnetic field and fragmentation of a young high-mass core (G11.92 MM2). We summarize our findings as follows:

1. In spite of having a relatively strong magnetic field  of 6.2 $\pm$ 3.5 mG, the combined effect of turbulence and the magnetic field cannot oppose the gravitational collapse of the core (virial parameter of 0.064 $\pm$ 0.028 and mass-to-magnetic flux ratio of 18 $\pm$ 9). In addition to being subvirialized and magnetically supercritical, the MM2 core is being fed with gas from its host filament at a rate of 1.9--5.6 \x 10$^{-4}$ \Msun\ yr$^{-1}$. Considering that the MM2 core is very young and can be used  as a proxy for the properties that the core had when it was prestellar, the formation scenario drawn in these observations contradicts the core accretion model and supports clump-fed scenarios. 

2. Despite of having only a minor contribution to the total energy budget at 1000 au scales (core scale), the magnetic field seems to be more important at a few 100s au scales influencing the fragmentation of MM2 and possibly shaping the properties of the binary system. Based on the analysis of energy ratios and a fragmentation criterion proposed from numerical simulations, the MM2 core could fragment following core fragmentation, but this fragmentation should be limited. Comparing the binary properties (mass ratio of 1.14 and separation of 505 au) and the MM2 core properties ($\mathcal{M}$ of 2.30 $\pm$ 0.03, $\mathcal{M_{\rm A}}$ of 1.6 $\pm$ 0.7, and $\lambda$ of 18 $\pm$ 9) with radiation-magnetohydrodynamical simulations, we conclude that we cannot rule out that the binary system could have been formed by disk fragmentation under the influence of super-Alfvenic, supersonic (or sonic) turbulence.  

\begin{acknowledgments}
P.S. was partially supported by a Grant-in-Aid for Scientific Research (KAKENHI Number JP22H01271 and JP23H01221) of JSPS. PS was supported by Yoshinori Ohsumi Fund (Yoshinori Ohsumi Award for Fundamental Research). P.C.C. was supported by the NAOJ Research Coordination Committee, NINS (NAOJ-RCC-2202-0401). The Green Bank Observatory is a facility of the National Science Foundation operated under cooperative agreement by Associated Universities, Inc. J.M.G. and P. Saha acknowledge support by the grant PID2020-117710GB-I00 (MCI-AEI-FEDER, UE). This work is also partially supported by the program Unidad de Excelencia Maria de Maeztu CEX2020-001058-M. P.S. was partially supported by a Grant-in-Aid for Scientific Research (KAKENHI numbers JP22H01271 and JP24K17100) of the Japan Society for the Promotion of Science (JSPS). M.T.B. acknowledges financial support through the INAF Large Grant {\it The role of MAGnetic fields in MAssive star formation}  (MAGMA). Y.C. was partially supported  by a Grant-in-Aid for Scientific Research (KAKENHI  number JP24K17103) of the JSPS. X.L.  acknowledges support from the National Key R\&D Program of China (No.\ 2022YFA1603101), the Strategic Priority Research Program of the Chinese Academy of Sciences (CAS) Grant No.\ XDB0800300, the National Natural Science Foundation of China (NSFC) through grant Nos.\ 12273090 and 12322305, the Natural Science Foundation of Shanghai (No.\ 23ZR1482100), and the CAS ``Light of West China'' Program No.\ xbzg-zdsys-202212. CJC acknowledges support from the STFC (grant ST/Y002229/1). K.P. is a Royal Society University Research Fellow, supported by grant number URF\textbackslash R1\textbackslash 211322. J.K. is supported by the Royal Society under grant number RF\textbackslash ERE\textbackslash 231132, as part of project URF\textbackslash R1\textbackslash 211322. L.A.Z. acknowledges financial support from CONACyT-280775, UNAM-PAPIIT IN110618, and IN112323 grants, M\'exico. 
CE acknowledges the financial support from the grant RJF/2020/000071 as a part of the Ramanujan Fellowship awarded by the Science and Engineering Research Board (SERB). 
Data analysis was in part carried out on the Multi-wavelength Data Analysis System operated by the Astronomy Data Center (ADC), National Astronomical Observatory of Japan. This paper makes use of the following ALMA data: ADS/JAO.ALMA\#2017.1.00101.S, ADS/JAO.ALMA\#2018.1.00105.S, ADS/JAO.ALMA\# 2016.1.01036.S , and ADS/JAO.ALMA\#2017.1.00237.S. ALMA is a partnership of ESO (representing its member states), NSF (USA) and NINS (Japan), together with NRC (Canada), MOST and ASIAA (Taiwan), and KASI (Republic of Korea), in cooperation with the Republic of Chile. The Joint ALMA Observatory is operated by ESO, AUI/NRAO and NAOJ.
\facilities{ALMA}
\software{CASA \citep[v5.1.1, 5.5; ][]{McMullin07}}
\end{acknowledgments}

%






\appendix

\section{Core properties excluding the magnetic field}

\subsection{Properties from Dust Continuum Emission} 
\label{sec:dust-app}

Sanhueza et al. (2025, in prep.) apply the dendrogram technique implemented in the Astrodendro Python package\footnote{https://dendrograms.readthedocs.io/en/stable/} \citep{Rosolowsky08} to define core properties. The input parameters used for the dendrogram are 5\,$\sigma$ as the minimum threshold for a leaf (core) detection, 1\,$\sigma$ as the minimum significance to separate leaves, and the minimum size for the definition of a leaf as the number of pixels equal to those contained in half of the synthesized beam. Among the many outputs from Astrodendro, we obtain a flux density (primary beam corrected) of 163 mJy and a radius ($R$) defined as half of the geometric mean of the FWHM of 0\farcs37 (1250 au) at 1.2 mm wavelength. 

The total gas mass is calculated from the dust continuum, assuming optically thin emission, as 
\begin{equation}
    M = \mathbb{R}~\frac{F_\nu D^2}{\kappa_\nu B_\nu (T)}~,
\label{eqn-dust-mass}
\end{equation}
where $\mathbb{R}$ is the gas-to-dust mass ratio, $F_\nu$ is the source flux density, 
$D$ is the distance to the source, $\kappa_\nu$ is the dust opacity per gram of dust, and $B_\nu$ is
 the Planck function at the dust temperature $T$. Assuming a gas-to-dust mass ratio of 100, dust opacity of 1.03 cm$^2$ g$^{-1}$ \citep[interpolated to 1.2 mm assuming $\beta$ = 1.6;][]{OH94}, and a temperature of 20 K \citep{Cyganowski14}, we  calculate a total mass of 31 $\pm$ 13 \Msun. 

 \cite{Cyganowski14} derive a dust temperature between 17-19 K for the  MM2 core using the (sub)millimeter spectral energy distribution and we adopt here an upper limit of 20 K. The uncertainty in the mass is dominated by the uncertainty from the gas-to-dust mass ratio and the dust opacity with 23\% and 28\% its respective values \citep{Sanhueza17}. The uncertainty for both the ALMA band 6 flux\footnote{https://almascience.nao.ac.jp/documents-and-tools/cycle11/alma-technical-handbook} and the parallax distance is on the order of 10\% \citep{Sato14}.  
 The number density, defined as $n$(H$_2$) = $M$/(Volume \x\ $\mu_{\rm H_2}m_{\rm H}$) with $\mu_{\rm H_2}$ being the molecular weight per hydrogen molecule and $m_{\rm H}$ the hydrogen mass is calculated assuming a spherical core of radius 1250 au. Assuming $\mu_{\rm H_2}=2.8$, we obtain a $n({\rm H}_2)$ of 4.8 $\pm$ 2.5 \x\ $10^8$~cm$^{-3}$. 

To estimate the accretion flows, the mass was calculated following the prescription above, assuming a dust temperature of 20 K. The flux for the blue-shifted and red-shifted flows is 21.5 and 13.3
mJy, resulting in 4.0 and 2.5 \Msun, respectively.

\subsection{Dynamics and Energetics} \label{sec:kin}

The observed line width at FWHM of the H$^{13}$CO$^+$ ($V_\mathrm{H^{13}CO^+}$) averaged inside the MM2 core is  1.44 $\pm$ 0.02 \kms, which results in a velocity dispersion $\sigma_\mathrm{H^{13}CO^+}$ of 0.61 $\pm$ 0.01 \kms\ ($V_\mathrm{H^{13}CO^+}$ = 1.44 \kms\ = $2\sqrt{2 \ln 2}\, \sigma_\mathrm{H^{13}CO^+}$). 

The total gas velocity dispersion is given by $\sigma_\mathrm{tot} = \sqrt{\sigma^2_\mathrm{th}+\sigma^2_\mathrm{nt}}$, in which the thermal velocity dispersion ($\sigma_\mathrm{th}$) and the non-thermal velocity dispersion ($\sigma_\mathrm{nt}$) are 

\begin{equation}
    \sigma^2_\mathrm{th} = \frac{k_{\rm B}T}{\mu_\mathrm{p} m_\mathrm{H}}
\end{equation}
and
\begin{equation}
    \sigma^2_\mathrm{nt} = \sigma^2_\mathrm{H^{13}CO^+} - \frac{k_{\rm B}T}{m_\mathrm{H^{13}CO^+}},
\end{equation}
respectively. $\mu_\mathrm{p}=2.33$ is the mean molecular weight per free particle considering H, He, and a negligible admixture of metals and $m_\mathrm{H^{13}CO^+}$ is the molecular mass of the H$^{13}$CO$^+$ equal to $30\,m_H$. 
Assuming that the non-thermal component is independent of the molecular tracer used, we can obtain a $\sigma_{\rm{tot}}$ of 0.66 $\pm$ 0.01 \kms. The value of $\sigma_{\rm{th}}$ is 0.26 \kms\ at 20 K and the sonic Mach number ($\mathcal{M}=\sigma_{\rm{nt}} / \sigma_{\rm{th}}$ = 0.61/0.26) is then 2.30 $\pm$ 0.03. 

The dynamical state of cores is generally assessed by using the virial theorem. The virial parameter ($\alpha_{\rm vir}$) is defined as the ratio between the virial mass ($M_{\rm vir}$) and the total mass (typically determined from dust continuum emission). A virial parameter of unity corresponds to equilibrium, $\alpha_{\rm vir} < 1$ implies gravitational collapse, $\alpha_{\rm vir} > 1$ means the core will  disperse. The most common virial analysis includes only gravity ($E_{\rm G}$) and the kinetic energy ($E_{\rm K}$, turbulence and thermal energy, neglecting rotation)
\begin{equation}
E_{\rm G} = -\frac{GM^2}{R}\left(\frac{3 - n}{5 - 2n} \right)~~~~~~{\rm and}~~~~~~E_{\rm K} = \frac{3}{2}M\sigma_{\rm tot}^2~.
\label{eq-Energy_1}
\end{equation}

The virial parameter can be expressed as
\begin{equation}
    \alpha_{\rm vir}=\frac{M_{\rm vir}}{M}=3\left(\frac{5-2n}{3-n}\right)\frac{R\sigma_{\rm tot}^2}{GM}~,
\label{eq-virial_1}
\end{equation}
resulting in $\alpha_{\rm vir}=0.060$ $\pm$ 0.026 for a centrally peaked density profile ($\rho(R) \propto R^{-n}$; $n=2$), indicating that turbulence alone cannot provide enough support against gravitational collapse.

According to \cite{Tsuribe99}, a core will fragment if the following two conditions are met: $\alpha' \beta' < 0.12$ and $\alpha' < 0.5$, in which $\alpha'$ ($=E_{\rm th}/E_{\rm G}$) is the  ratio of the thermal to gravitational energy and 
$\beta'$ ($=E_{\rm rot}/E_{\rm G}$) is the ratio of the rotational to gravitational energy. To obtain the thermal energy ($E_{\rm th}$), we replace $\sigma_{\rm tot}$ by $\sigma_{th}$ in the right side of Equation~\ref{eq-Energy_1}. Following \cite{Sanhueza21}, the rotational energy ($E_{\rm rot}$) is given by 
\begin{equation}
E_{\rm rot} = \frac{1}{3}Mv_{\rm rot}^2\left(\frac{3 - n}{5 - n} \right)~,
\label{rot-eq}
\end{equation}
in which $v_{rot}$ is the rotational velocity. Because the core MM2 shows no clear rotation at the current spectral resolution, we adopt the spectral resolution as an upper limit for the rotation, i.e., $v_{\rm rot}=0.56$ \kms. 
 For the MM2 core, we obtain $\alpha' \beta' = 8 \pm 7 \times 10^{-6}$ and $\alpha' = 5 \pm 2 \times 10^{-3}$,  indicating that the binary system could be formed by fragmentation at the core scale. 

To assess whether the conclusions of this work are affected by the temperature adopted, we have calculated the core properties also assuming a dust temperature of 50 K. The values are compared with those obtained at 20 K in Table~\ref{tab:no_B}. 

\begin{deluxetable}{cccccccccccc}[h]
\label{tab:no_B}
\tabletypesize{\footnotesize}
\tablecaption{Core Properties at Different Dust Temperatures Excluding the Magnetic Field}
\tablewidth{0pt}
\tablehead{
\colhead{$T$}  & \colhead{$M$} & \colhead{$n({\rm H}_2)$ } & \colhead{$\sigma_\mathrm{th}$} & \colhead{$\sigma_{\rm{nt}}$} & \colhead{$\sigma_{\rm{tot}}$} & \colhead{$\mathcal{M}$} & \colhead{$E_{\rm G}$} & \colhead{$E_{\rm K}$} & \colhead{$E_{\rm th}$} & \colhead{$E_{\rm rot}$} & \colhead{$\alpha_{\rm vir}$}\\ 
\colhead{(K)} & \colhead{(\Msun)} & \colhead{(cm$^{-3}$)}  & \colhead{(\kms)} & \colhead{(\kms)} & \colhead{(\kms)} & & \colhead{(ergs)} & \colhead{(ergs)} & \colhead{(ergs)}  & \colhead{(ergs)} & \colhead{}\\
\colhead{} & \colhead{} & \colhead{$\x10^8$}  & \colhead{} & \colhead{} & \colhead{} & & \colhead{$\x10^{46}$} & \colhead{$\x10^{44}$} & \colhead{$\x10^{43}$}  & \colhead{$\x10^{43}$} & \colhead{}
}
\startdata
20 & $31\pm13$  & $4.8\pm2.5$ & 0.26 & $0.61\pm0.01$ & $0.66\pm0.01$ & $2.30\pm0.03$ & $1.3\pm1.1$  & $4.0\pm1.7$ & $6.4\pm2.7$ & $2.1\pm0.9$  & $0.060\pm0.026$\\
50 & $10\pm4$  & $1.6\pm0.8$ & 0.42 & $0.60\pm0.01$ & $0.73\pm0.01$ & $1.44\pm0.02$ & $0.14\pm0.12$ & $1.6\pm0.7$ & $5.2\pm2.2$ & $0.7\pm0.3$ & $0.22\pm0.10$ \\
\enddata
\end{deluxetable}

\section{Magnetic Field Properties} \label{sec:bfield-app1}

Following the procedure of \citet{Liu24}, we use the ADF method \citep{Houde16} to estimate the plane-of-sky total (ordered+turbulent) magnetic field strength as 
\begin{equation}
B = 0.21 \sqrt{4\pi \rho }~\sigma_{\rm v}\left(\frac{\langle B_\mathrm{t}^2 \rangle}{\langle B^2 \rangle}\right)^{-0.5},
\end{equation}
where $\rho$ is the gas density, $\sigma_{\mathrm{v}}$ is the turbulent velocity dispersion (assumed to be $\sigma_{\rm nt}=0.61$ \kms), and $(\langle B_\mathrm{t}^2 \rangle/\langle B^2 \rangle)^{0.5}$ is the turbulent-to-total magnetic field strength ratio without correction for LOS integration. Because the ADF may not correctly account for the LOS signal integration in high-density regions, we additionally adopt a numerical correction factor of 0.21 to account for this effect \citep{Liu21}. The turbulent-to-total field ratio
$(\langle B_\mathrm{t}^2 \rangle/\langle B^2 \rangle)^{0.5}$ is derived by fitting the ADF 
\begin{equation}
1 - \langle \cos \lbrack \Delta \Phi (l)\rbrack \rangle =  a_{2} l^{2} + \frac{\langle B_\mathrm{t}^2 \rangle}{\langle B^2 \rangle} C \nonumber \\
\times \Bigg\{ \frac{1}{C_1} \left[ 1 - e^{-l^2/2(l_\delta^2+2l_{W1}^2)}\right] \nonumber \\+ \frac{1}{C_2} \left[ 1 - e^{-l^2/2(l_\delta^2+2l_{W2}^2)}\right] \nonumber \\- \frac{2}{C_{12}} \left[ 1 - e^{-l^2/2(l_\delta^2+l_{W1}^2+l_{W2}^2)}\right] \Bigg\},
\end{equation}
where $\Delta \Phi (l)$ is the angular difference of two position angles separated by $l$, $l_{W1}$ is the ALMA beam size divided by 
$2\sqrt{2 \ln 2}$, $l_{W2}$ is the maximum recoverable scale of ALMA divided by $2\sqrt{2 \ln 2}$, $a_2 l^2$ is the second-order term of the Taylor expansion for the ordered field, and $l_\delta$ is the turbulent correlation length. The coefficients are given by 
\begin{equation}
C_1 = \frac{(l_\delta^2 + 2l_{W1}^2)}{\sqrt{2\pi}l_\delta^3},
\end{equation}
\begin{equation}
C_2 = \frac{(l_\delta^2 + 2l_{W2}^2)}{\sqrt{2\pi}l_\delta^3},
\end{equation}
\begin{equation}
C_{12} = \frac{(l_\delta^2 + l_{W1}^2 + l_{W2}^2)}{\sqrt{2\pi}l_\delta^3},
\end{equation}
\begin{equation}
C = \left( \frac{1}{C_1} + \frac{1}{C_2} - \frac{2}{C_{12}} \right)^{-1}.
\end{equation}
Note that our equations are slightly different from the original ones in \citet{Houde16} because the effect of LOS signal integration is considered differently. By fitting the ADF of MM2 with approaches similar to \citet{Liu20, Liu24}, we obtain $(\langle B_\mathrm{t}^2 \rangle/\langle B^2 \rangle)^{0.5} = 0.43$, with a statistical uncertainty of 45\% its value when the numerical correction of 0.21 is used. The resulting plane-of-sky total magnetic field strength is $B=5.0\pm2.8$ mG. Note that the energy equipartition assumption of DCF may not be satisfied and the line-of-sight signal integration could be more significant in high-density regions \citep{Liu21}, so the estimated field strength may only be an upper limit. Adopting the statistical relation $B_{3D} \sim B \times 1.25$ \citep{Liu22b}, we obtain the 3D total field strength of $B_{3D}=6.2\pm3.5$ mG. The Alfv\'en speed, given by $\sigma_{\rm A} = B/\sqrt{4\pi\rho}$, is $0.37\pm0.23$ \kms, resulting in an Alfvenic Mach number ($\mathcal{M_{\rm A}}=\sigma_{\rm{nt}} / \sigma_{\rm{A}}=0.60/0.37$) of $1.6\pm0.7$. 

The relative importance between magnetic field and gravity is usually characterized by the normalized mass-to-magnetic flux ratio \citep{Crutcher2004}. We calculate the normalized mass-to-magnetic flux ratio as in \cite{Liu22a}
\begin{equation}
    \lambda = \frac{(M/\Phi_B)}{(M/\Phi_B)_{\rm cr}} = 2 \pi G^{1/2}\left[\frac{3}{2}\left(\frac{3-n}{5-2n}\right)\right]^{1/2}\frac{M}{\Phi_B}~,
\label{lambda}    
\end{equation}
where $\Phi_B = B \pi R^2$ is the magnetic flux. For $n=2$, we obtain $\lambda = 18\pm13$, which suggests a magnetically supercritical state where gravity dominates the magnetic field. 

The virial parameter considering support from both the turbulence and the magnetic field can be written as \citep{Liu24}
\begin{equation}\label{eq:virialtotalE}
\alpha_{\mathrm{vir,B}} = \frac{2E_{\mathrm{K}} + E_\mathrm{B}}{\vert E_{\mathrm{G}}\vert},
\end{equation}
where the magnetic energy ($E_{\rm B}$) is given by
\begin{equation}
E_{\rm B} = \frac{1}{8\pi}B^2V = \frac{1}{6}B^2R^3~.
\end{equation}
 For MM2, we obtain $\alpha_{\mathrm{vir,B}} = 0.064~\pm~0.028$, which is only slightly larger than the kinetic virial parameter ($\alpha_{\mathrm{vir}}=0.060~\pm~0.026$), but indistinguishable within the uncertainties. The small virial parameter indicates a dynamical collapsing state of MM2, far from equilibrium. 

Table~\ref{tab:B} shows the core properties including the magnetic field for a dust temperature of 50 K. 

\begin{deluxetable}{ccccccc}[h]
\label{tab:B}
\tabletypesize{\footnotesize}
\tablecaption{Magnetic Field Properties at Different Dust Temperatures}
\tablewidth{0pt}
\tablehead{
\colhead{$T$} & \colhead{$B_{3D}$}  & \colhead{$\sigma_{\rm A}$} & \colhead{$\mathcal{M_{\rm A}}$} & \colhead{$\lambda$} & \colhead{$E_{\rm B}$} & \colhead{$\alpha_{\mathrm{vir,B}}$} \\ 
\colhead{(K)} & \colhead{(mG)} & \colhead{(\kms)} & \colhead{(cm$^{-3}$)}  & \colhead{(\kms)} & \colhead{(erg)}
}
\startdata
20 & $6.2\pm3.5$ & $0.37\pm0.23$ & $1.6\pm0.7$ & $18\pm13$ & $4.2\pm4.1$ \x\ $10^{43}$ & $0.064\pm0.028$\\
50 & $3.5\pm2.0$ & $0.37\pm0.23$ & $1.6\pm0.7$ & $10\pm5$ & $1.4\pm1.3$ \x\ $10^{43}$ & $0.23\pm0.10$\\
\enddata
\end{deluxetable}


\bibliography{references}{}
\bibliographystyle{aasjournal}




\end{document}